# Compositionally Complex Perovskite Oxides for Solar Thermochemical Water Splitting

Dawei Zhang [a,+], Héctor A. De Santiago [b,+], Boyuan Xu [c], Cijie Liu [b], Jamie Trindell [d], Wei Li [b,*], Jiyun Park [e], Mark A. Rodriguez [f], Eric N. Coker [f], Josh Sugar [d], Anthony McDaniel [d], Stephan Lany [g], Liang Ma [b,h], Yi Wang [b], Gregory Collins [b], Hanchen Tian [b], Wenyuan Li [i], Yue Qi [e,*], Xingbo Liu [b,*], and Jian Luo [a,j,*]

[a] Program of Materials Science and Engineering, University of California San Diego, La Jolla, CA 92093, USA
[b] Department of Mechanical and Aerospace Engineering, Benjamin M. Statler College of Engineering and Mineral Resources, West Virginia University, Morgantown, WV 26506, USA
[c] Department of Physics, Brown University, Providence, Rhode Island 02912, USA
[d] Sandia National Laboratories, Livermore, CA 94551, USA
[e] School of Engineering, Brown University, Providence, Rhode Island 02912, USA
[f] Sandia National Laboratories, Albuquerque, NM 87185, USA
[g] Materials Science Center, National Renewable Energy Laboratory, Golden, Colorado 80401, USA
[h] School of Materials Science and Engineering, Hebei University of Engineering, Handan, Hebei Province 056038, China
[i] Department of Chemical and Biomedical Engineering, Benjamin M. Statler College of Engineering and Mineral Resources, West Virginia University, Morgantown, WV 26506, USA
[j] Department of NanoEngineering, University of California San Diego, La Jolla, CA 92093, USA

## Abstract

Solar thermochemical hydrogen generation (STCH) is a promising approach for eco-friendly $H_2$ production, but conventional STCH redox compounds often suffer from thermodynamic and kinetic limitations with limited tunability. Expanding from the nascent high-entropy ceramics field, this study explores a new class of compositionally complex perovskite oxides $(La_{0.8}Sr_{0.2})(Mn_{(1-x)/3}Fe_{(1-x)/3}Co_xAl_{(1-x)/3})O_3$ for STCH. *In situ* X-ray diffraction demonstrates the phase stability during redox cycling and *in situ* X-ray photoelectron spectroscopy shows preferential redox of Co. The extent of reduction increases, but the intrinsic kinetics decreases, with increased Co content. Consequently, $(La_{0.8}Sr_{0.2})(Mn_{0.2}Fe_{0.2}Co_{0.4}Al_{0.2})O_{3-\delta}$ achieves an optimal balance between the thermodynamics and kinetics properties. The combination of a moderate enthalpy of reduction, high entropy of reduction, and preferable surface oxygen exchange kinetics enables a maximum $H_2$ yield of $395 \pm 11$ μmol g$^{-1}$ in a short 1-hour redox duration. Entropy stabilization expectedly contributes to the structure stability during redox without phase transformation, which enables an exceptional STCH stability for >50 cycles under harsh interrupted conditions. The underlying redox mechanism is further elucidated by the density functional theory based parallel Monte Carlo computation, which represents a new computation paradigm first established here. This study suggests a new class of non-equimolar compositionally complex ceramics for STCH and chemical looping.

---

[+] These authors contributed equally to this work.
[*] Corresponding authors: Jian Luo (jluo@alum.mit.edu), Xingbo Liu (Xingbo.Liu@mail.wvu.edu), Yue Qi (yue_qi1@brown.edu), and Wei Li (wei.li@mail.wvu.edu)

**Table of Content (ToC)**

A new class of compositionally complex perovskite oxides $(La_{0.8}Sr_{0.2})(Mn_{(1-x)/3}Fe_{(1-x)/3}Co_xAl_{(1-x)/3})O_3$ were discovered for two-step redox solar thermochemical hydrogen production by expanding the emerging field of high-entropy ceramics to non-equimolar compositional complex ceramics. The *in-situ* characterization and density functional theory based parallel Monte Carlo computation elucidated the redox sequence and underlying mechanism.

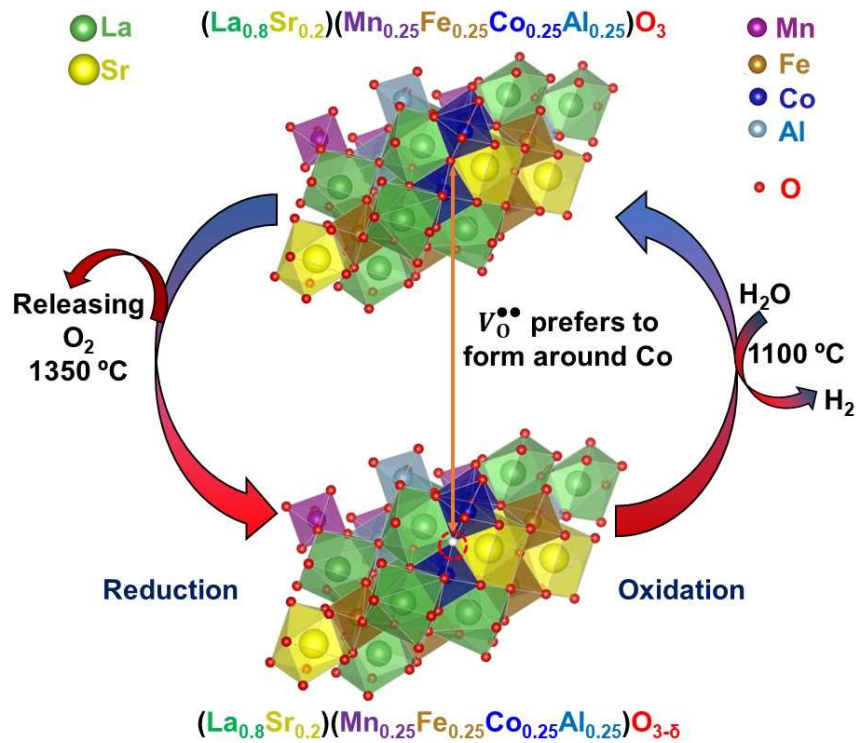

## 1. Introduction

Hydrogen has critical applications in modern industries and energy systems for future hydrogen economy.[1–5] To realize sustainable sunlight-driven water splitting and store the intermittent solar energy, two-step solar thermochemical hydrogen production (STCH) has attracted considerable attention due to its capability of utilizing the full solar spectrum energy and potential to generate industrial-scale quantities of solar fuels.[6–11] The two-step STCH process involves an endothermic reduction step at a high temperature (≥1200 °C) under low $P_{O_2}$ with a suitable non-stoichiometric redox metal oxide to release oxygen and a subsequent oxidation step via flowing steam to the reduced oxide at a relatively lower temperature (~800-1100 °C). The most investigated non-stoichiometric metal oxide materials for STCH are undoped and doped ceria materials due to their structural stability and fast redox kinetics.[12–15] However, ceria suffers from the requirement of extremely high reduction temperature (>1500 °C) to reach a small extent of reduction ($\Delta\delta \approx$ 0.03-0.06), challenging the design of suitable reactors.[14,16] The perovskite oxides ($ABO_{3-\delta}$) are considered as a promising alternative candidate to ceria for STCH, as they have favorable merits including structural stability under large non-stoichiometry redox swing, tunable defect chemistry with A/B site doping, and a large compositional space.[17,18] For instance, $Ca(Ti_{0.5}Mn_{0.5})O_3$[19] and $(La_{1-x}Sr_x)(Mn_{1-y}Al_y)O_3$[20] perovskites have been investigated for STCH. Even though perovskites can have much larger redox capacities ($\Delta\delta$ > 0.15), longer reaction time (total dwell time for reduction and oxidation steps > 2 h) is usually required to reach a fair $H_2$ yield in the reports.[21–23] Understanding of the tradeoff between thermodynamic and kinetics properties is needed to optimize the perovskite composition to achieve high $H_2$ yield within a short time (≤ 1h).

High-entropy ceramics (HECs) have been synthesized in several material families including rocksalt,[24,25] perovskite,[26] and fluorite[27–29] oxides, as well as other non-oxides such as borides,[30] carbides[31] and silicides.[32] Notably, an equimolar poly-cation $(Fe_{0.25}Mg_{0.25}Co_{0.25}Ni_{0.25})O_x$ was reported for STCH undergoing reversible phase transformation between the spinel and rocksalt structure for 10 cycles.[33] However, the repetitive phase transformation is not desirable for long-term cycling stability, reversibility, and oxygen exchange kinetics due to its larger energy barrier. Fe was proposed as the redox active element, although the *ex-situ* characterization in that work may not precisely track the instantaneous redox change and the redox center valence may subject to change in the quenching/cooling process. It remains unclear why a particular cation dominates the redox chemistry and what factors govern the priority of redox behavior in the HECs during the STCH process. Recently, compositionally complex ceramics (CCCs) have been introduced to broaden the field of HECs by including non-equimolar compositions that reduce the configurational entropy but allow more engineering space to improve the properties.[34,35] CCCs can provide additional tunability of the physical properties (*e.g.* thermomechanical properties in compositionally complex fluorite-based oxides).[36,37] Here, we hypothesize that a vast compositional space in compositionally complex pervoskite oxides (CCPOs) will enable us to create a new class of stable STCH materials, where non-equimolar compositional designs can be leveraged to balance the thermodynamic and kinetic properties.

Herein, we designed a new class of medium- to high-entropy CCPOs for two-step STCH with the chemical formula $(La_{0.8}Sr_{0.2})(Mn_{(1-x)/3}Fe_{(1-x)/3}Co_xAl_{(1-x)/3})O_3$, denoted as "LS_MFC$_x$A" for brevity (**Table 1**). Unlike $(Fe_{0.25}Mg_{0.25}Co_{0.25}Ni_{0.25})O_x$,[33] and $BaCe_xMn_{1-x}O_3$,[38] this LS_MFC$_x$A series of CCPOs did not show any phase transformation during the STCH redox cycling for $x < 0.61$. In this series, the thermodynamic and intrinsic kinetics properties showed different correlations with the Co content. LS_MFC$_{0.4}$A or $(La_{0.8}Sr_{0.2})(Mn_{0.2}Fe_{0.2}Co_{0.4}Al_{0.2})O_3$ ($x = 0.4$), showed the best balance between the thermodynamics and kinetics properties, rendering a high hydrogen yield of $395 \pm 11$ μmol g$^{-1}$ within a short 1 h duration at an optimized STCH condition. *In situ* characterization indicated that Co dominated the redox. Furthermore, Monte Carlo (MC) sampling based on density functional theory (DFT) was applied in this field, for the first time to our knowledge, to demonstrate that oxygen vacancies prefer to form on the Co octahedron position among all B-site metal octahedron positions and the valence change of Co is the most obvious. Furthermore, we found that the Co-O bond in the CCPO is substantially weakened in comparison to both Mn-O, Al-O and Fe-O bonds and their respective simple perovskites, rationalizing the experimental observations. Moreover, LS_MFC$_{0.4}$A exhibited exceptional structural stability and cycling durability even after 51 cycles under very harsh interrupted cycling conditions involving startup heating and shutdown cooling, which simulates the real-world day-night cycle conditions for STCH. This work opens up a new direction to explore novel CCCs for STCH and provides a new computation paradigm for mechanistic understanding of redox chemistries of STCH materials.

**Table 1**. Compositions of CCPOs and abbreviations.

| Sample Abbreviations | Nominal Composition |
| --- | --- |
| LS_MFC$_{0.16}$A | $(La_{0.8}Sr_{0.2})(Mn_{0.28}Fe_{0.28}Co_{0.16}Al_{0.28})O_{3-\delta}$ |
| LS_MFC$_{0.2}$A | $(La_{0.8}Sr_{0.2})(Mn_{0.267}Fe_{0.267}Co_{0.20}Al_{0.267})O_{3-\delta}$ |
| LS_MFC$_{0.25}$A | $(La_{0.8}Sr_{0.2})(Mn_{0.25}Fe_{0.25}Co_{0.25}Al_{0.25})O_{3-\delta}$ |
| LS_MFC$_{0.4}$A | $(La_{0.8}Sr_{0.2})(Mn_{0.2}Fe_{0.2}Co_{0.4}Al_{0.2})O_{3-\delta}$ |
| LS_MFC$_{0.52}$A | $(La_{0.8}Sr_{0.2})(Mn_{0.16}Fe_{0.16}Co_{0.52}Al_{0.16})O_{3-\delta}$ |
| LS_MFC$_{0.61}$A | $(La_{0.8}Sr_{0.2})(Mn_{0.13}Fe_{0.13}Co_{0.61}Al_{0.13})O_{3-\delta}$ |
| LS_MFC$_{0.79}$A | $(La_{0.8}Sr_{0.2})(Mn_{0.07}Fe_{0.07}Co_{0.79}Al_{0.07})O_{3-\delta}$ |
| LSC | $(La_{0.8}Sr_{0.2})CoO_3$ |

## 2. Results and Discussion

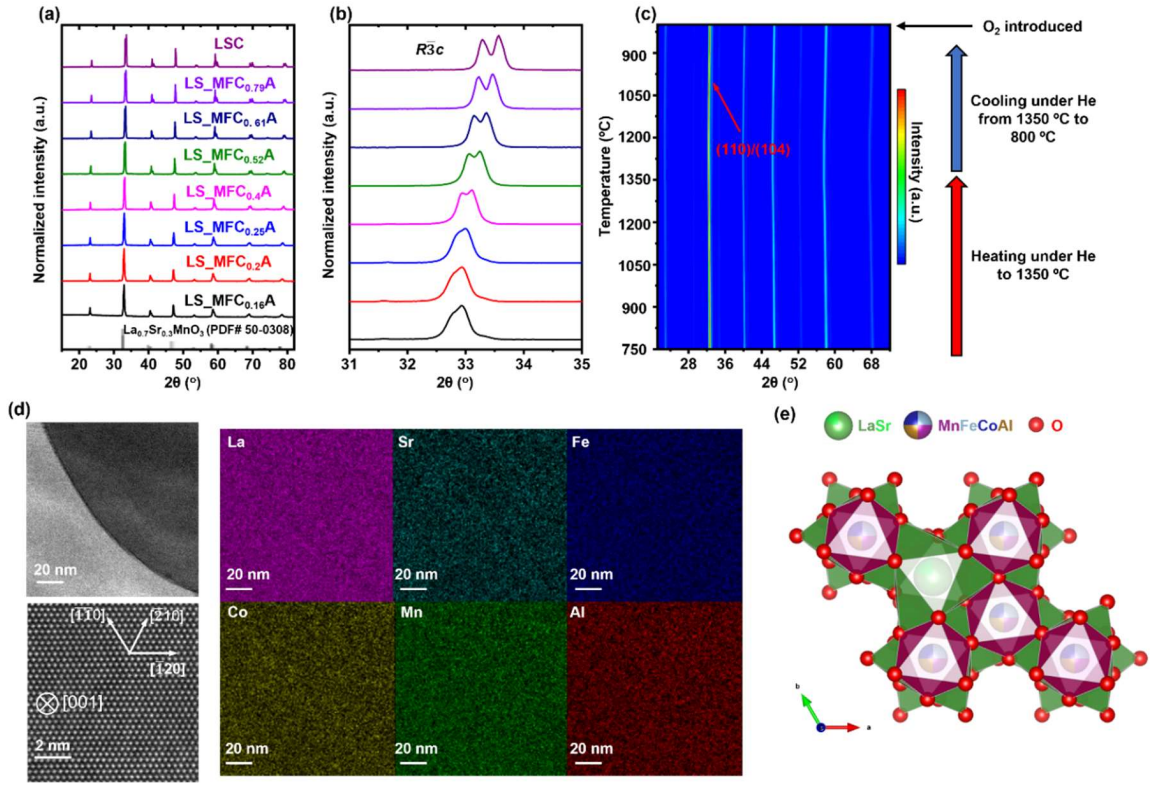

**Figure. 1 a)** XRD patterns of LS_MFC$_x$A ($x$ = 0.16, 0.2, 0.25, 0.4, 0.52, 0.61, 0.79, and 1). **b)** The splitting of (110) and (104) peaks showing the $R\bar{3}c$ rhombohedral structure. **c)** HT-XRD patterns of this LS_MFC$_{0.16}$A series. The sample was heated up from room temperature to 1350 °C under a He flow. There was no phase transformation or separation occurring in LS_MFC$_{0.16}$A under thermally reduced and oxidation conditions. **d)** STEM HAADF images of LS_MFC$_{0.25}$A (with a grain boundary) and its nanoscale EDS elemental maps. **e)** Schematic illustration of the corresponding crystal structure of LS_MFC$_{0.25}$A.

The (La$_{0.8}$Sr$_{0.2}$)(Mn$_{(1-x)/3}$Fe$_{(1-x)/3}$Co$_x$Al$_{(1-x)/3}$)O$_3$ (LS_MFC$_x$A) CCPOs were prepared by solid-state reactions. Table 1 shows the sample abbreviations and nominal compositions for this LS_MFC$_x$A series of CCPOs. The X-ray diffraction (XRD) pattern (**Figure 1a**) reveals that LS_MFC$_x$A adopted a rhombohedral structure ($R\bar{3}c$), with the peak splitting at 2θ of 32.8° (**Figure 1b**). Rietveld refinements confirm the $R\bar{3}c$ phase for all compositions by assuming random B site occupation of Mn, Fe, Co, and Al (**Figure S1** and **Table S1**, Supporting Information). As the Co content increases in LS_MFC$_x$A, the peak splitting becomes larger due to more severe lattice distortion and the diffraction peaks gradually shift to higher 2θ angle indicating the decrease of lattice parameters. To probe the structural stability of LS_MFC$_x$A under thermally reducing environments, LS_MFC$_{0.16}$A was selected to perform high-temperature XRD. **Figure 1c** displays that LS_MFC$_{0.16}$A showed structural stability at 1350 °C and maintained the $R\bar{3}c$ rhombohedral structure without phase transition under the reducing conditions, except for the slight peak shifts (lattice expansion) due to the thermal expansion and oxygen loss. After the $O_2$ uptake at 800 °C, the lattice was recovered to the initial state. Typical HT-XRD patterns were refined (**Figure S2**) showing the crystallographic structure was maintained during the thermal reduction and oxidation conditions. The atomic structure and elemental distribution of LS_MFC$_{0.25}$A were characterized by the scanning transmission electron microscopy (STEM) high-angle annular dark field (HAADF) imaging with the energy dispersive spectroscopy (EDS) (**Figure 1d**), which suggested a homogenous elemental distribution of cation elements. The XRD and EDS mapping results indicate a long-range random B site occupation.

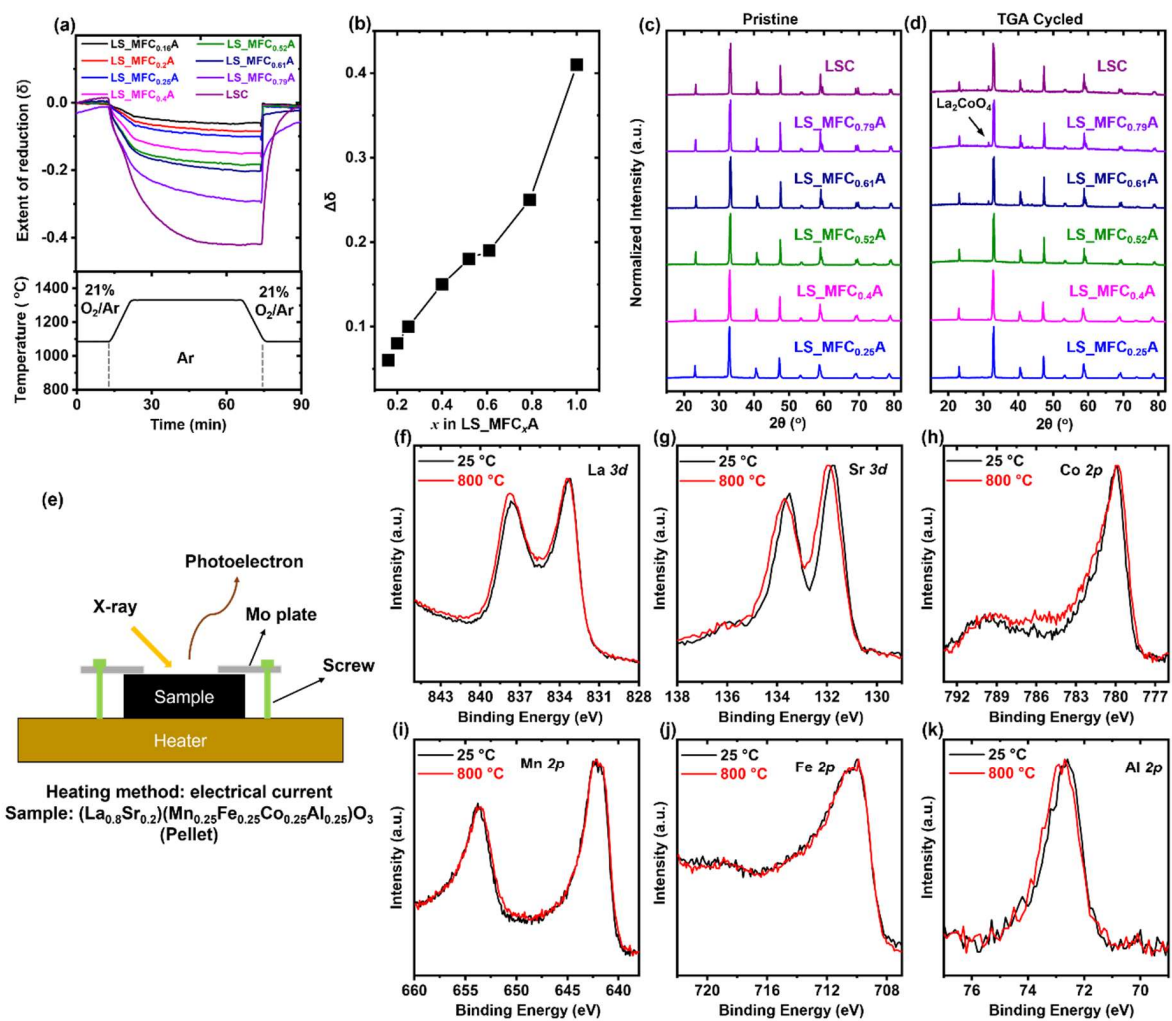

**Figure 2. a)** TPR tests of the LS_MFC$_x$A with $x$ ranging from 0.16 to 1 to evaluate the extent of reduction (formation of oxygen vacancies) at 1350 °C for 45 min in Ar and reoxidation at 1100 °C in 21% $O_2$ balanced with Ar. **b)** The correlation of $\Delta\delta$ with $x$ in LS_MFC$_x$A. Comparison of XRD patterns before **c)** and after **d)** TPR experiments. La$_2$CoO$_4$ secondary phase formed in LS_MFC$_x$A ($x > 0.52$). **e)** Schematic of the *in situ* XPS set-up and the normalized **f)** La *3d*, **g)** Sr *3d*, **h)** Co *2p*, **i)** Mn *2p*, **j)** Fe *2p* and **k)** Al *2p* peaks of LS_MFC$_{0.25}$A at 25 and 800 °C. All energy levels were calibrated by C *1s* at 284.8 eV.

To quantify the redox capability, LS_MFC$_x$A samples were tested by temperature-programmed reduction (TPR) in thermogravimetric analysis (TGA).[39] **Figure 2a** shows the TPR results for LS_MFC$_x$A specimens at a reduction temperature of 1350 °C for 45 min under Ar and oxidation temperature of 1100 °C for 45 min under 21% O$_2$ balanced with Ar. The reversible extent of reduction (Δδ) displays a decreasing trend with narrower temperature swing (**Figure S3**, Supporting Information). The TPR experiments show that Δδ monotonically increases with the molar ratio of Co in the B site of LS_MFC$_x$A (**Figure 2b**). The endmember of simple perovskite (La$_{0.8}$Sr$_{0.2}$)CoO$_3$ (LSC) can greatly be reduced with Δδ = 0.4 compared to LS_MFC$_{0.16}$A with the smallest Δδ = 0.06. When $x \leq 0.52$, the reoxidation occurred instantly and reduced samples were fully recovered under 21% O$_2$ environment. However, when $x$ was increased beyond ~0.61, the reoxidation underwent a sluggish recovery to the initial state. The slow reoxidation rate under a high $P_{O_2}$ of 21% indicates that the driving force for reoxidation can be a limited factor for these compositions (LS_MFC$_x$A, $x \geq 0.52$). The sluggish reoxidation process also implies a potential irreversible phase transformation or secondary phase segregation during the redox cycling. The XRD patterns for LS_MFC$_x$A before and after the TGA measurements were compared (**Figure 2c** and **d**). A Co-enriched Ruddlesden-Popper (RP) La$_2$CoO$_4$ secondary phase was formed in LS_MFC$_{0.61}$A, LS_MFC$_{0.79}$A and LSC which showed sluggish reoxidation kinetics. In contrast, LS_MFC$_x$A ($x \leq 0.52$) showed no detectable secondary phase in the XRD patterns and displayed rapid reoxidation. Therefore, the structural stability and redox reversibility of LS_MFC$_x$A ($x \leq 0.52$) make them suitable for STCH. Besides Co, we have also tuned other B site elements deviating from the equal-molar composition and found that Co had the largest influence on the extent of reduction (**Figure S4**, Supporting Information).

To further confirm the redox sequence of B-site elements in the LS_MFC$_x$A, the *in-situ* X-ray photoelectron spectroscopy (XPS) was conducted to study the redox behavior of all cations by heating LS_MFC$_{0.25}$A to 800 °C (XPS heating limit) under vacuum (**Figure 2e**). The high-resolution XPS spectra of all cations at 25 and 800 °C are shown in **Figure 2f-k**. LS_MFC$_{0.25}$A showed a fair extent of reduction at 800 °C (**Figure S5**, Supporting Information). For the A-site elements, the left peak area of La $3d_{5/2}$ slightly increased indicating pyrolysis of the surface absorbed La(OH)$_3$ species at 800 °C. Sr also exhibited a side peak evolution, which was assigned to slight Sr segregation that is widely found in solid oxide fuel cells.[40–42] For the B-site elements, Mn, Fe, and Co can be potentially redox active. Al $2p$ showed negligible shift due to the stable Al$^{3+}$ valence state. Co $2p$ showed an increased intensity and broadening of the side peak located at 787 eV, which is a Co$^{2+}$ $2p_{3/2}$ satellite peak. This satellite peak evolution suggests that Co$^{3+}$ is reduced to Co$^{2+}$ during the *in-situ* heating process, in agreement with previous reports.[43–45] In contrast, Fe $2p$ and Mn $2p$ had negligible changes upon heating. Meanwhile, the *ex-situ* STEM electron energy loss spectroscopy (EELS) of LS_MFC$_{0.4}$A demonstrates that only Co has obvious valence change (**Figure S6**, Supporting Information). Therefore, Co is likely the primary redox element, consistent with the correlation between Δδ and Co content in the TPR.

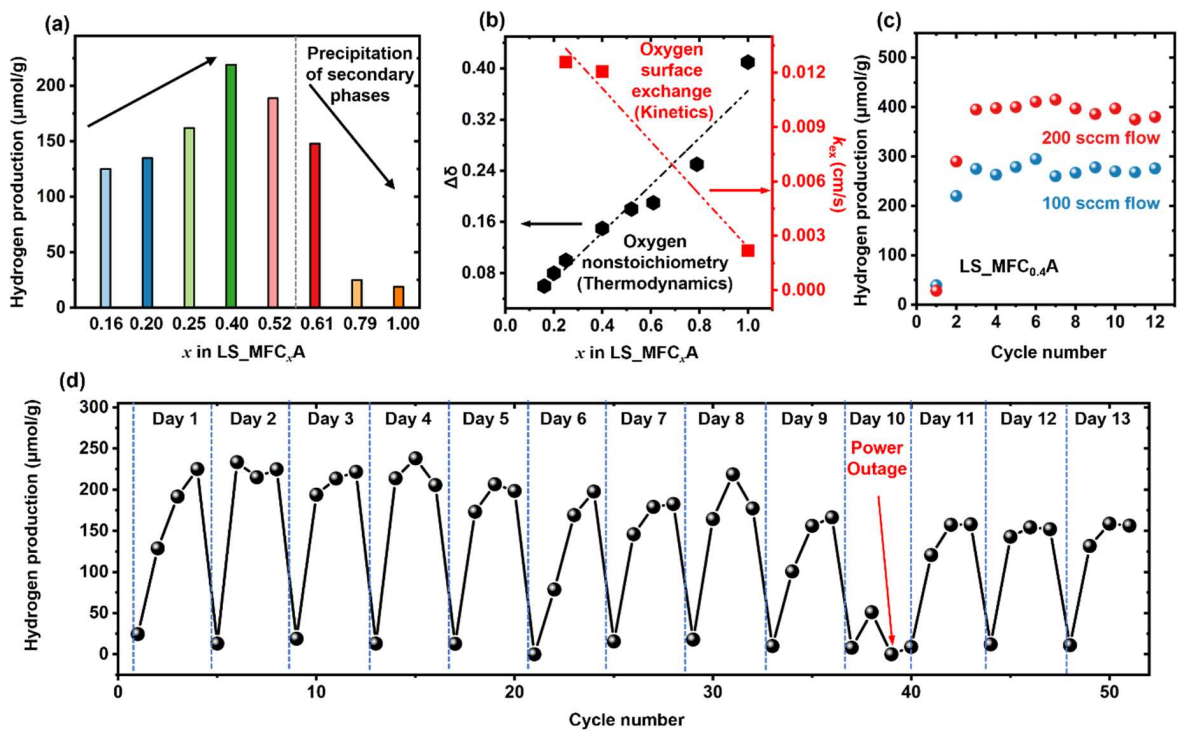

**Figure 3. a)** Average hydrogen production of two stabilized cycles for LS_MFC$_x$A. The reduction was conducted at 1350 °C ($T_{Re}$) in N$_2$ for 45 min and oxidation was performed at 1100 °C ($T_{Ox}$) in 40 vol% H$_2$O for 15 min with a gas flow rate of 100 sccm. H$_2$ production shows a volcano plot shape with respect to Co molar ratio in LS_MFC$_x$A. **b)** The dependences of Δδ (measured by TPR) and oxygen surface exchange coefficient $k_{ex}$ (measured by electrical conductivity relaxation) on the Co content in LS_MFC$_x$A. The thermodynamic (Δδ) and kinetic ($k_{ex}$) properties demonstrate opposite correlation with Co molar ratio in LS_MFC$_x$A. **c)** H$_2$ production stability of LS_MFC$_{0.4}$A under different flow rate during the uninterrupted cycling ($T_{Re}$ = 1350 °C for 30 min in N$_2$ and $T_{Ox}$ = 1100 °C for 30 min in 40 vol% steam). **d)** Long-term STCH cycling of LS_MFC$_{0.4}$A under harsh interrupted cycling conditions involving startup heating and shutdown cooling, at a STCH condition producing significant (but not maximized) H$_2$ yield. Specifically, the sample was heated in the morning and cooled down to room temperature at night and four cycles were performed per day (using the STCH condition: $T_{Re}$ = 1350 °C for 45 min in N$_2$ and $T_{Ox}$ = 1100 °C for 15 min in 40 vol% steam with a gas flow rate of 100 sccm). This simulates the real day-night cycle conditions for STCH than uninterrupted cycling. Background H$_2$ production from catalytic water thermolysis was subtracted from the total H$_2$ yield and only the H$_2$ production from the redox reaction was illustrated.

The STCH performances of LS_MFC$_x$A materials were investigated using a homemade flow reactor (**Figure S7**, Supporting Information). It is worth noting that the water thermolysis catalyzed by the alumina tube and samples could take place at 1100 °C in 40 vol% steam flow producing a small quantity of H$_2$, which has been recently reported.[46–48] Thus, the H$_2$ production from direct catalytic water splitting was measured (**Figure S8**, Supporting Information) and subtracted from the total H$_2$ yield to extract the true H$_2$ yield from the redox reaction. **Figure 3a** shows H$_2$ yields of various LS_MFC$_x$A under the given conditions. The hydrogen yield increased from LS_MFC$_{0.16}$A to LS_MFC$_{0.4}$A and then rapidly fell after $x$ = 0.61. The relationship of H$_2$ yield with the Co content has a volcano shape with the peak H$_2$ yield achieved by LS_MFC$_{0.4}$A, similar to the observed trend in La(Ga$_{1-y}$Co$_y$)O$_3$.[21] Under this condition, LS_MFC$_{0.4}$A produced the highest H$_2$ yield (219 μmol/g) although it had a medium Δδ. In contrast, the simple perovskite LSC showed the minimal H$_2$ yield (24 μmol/g) despite its largest Δδ. The molar ratio of H$_2$ to O$_2$ produced in the oxidation and reduction steps, respectively, is 2:1 (**Figure S9**), suggesting the water splitting is realized by the redox reaction.

The volcano shaped plot of hydrogen production *vs.* Co molar ratio may relate to the trade-off between the thermodynamic and kinetic properties as shown in **Figure 3b**. The intrinsic kinetic properties of the LS_MFC$_x$A series were examined by using the electrical conductivity relaxation (ECR) technique, which can quantify the oxygen surface reaction coefficient $k_{ex}$ (*i.e.*, the proportionality between the rate of oxygen incorporation and deviation of surface oxygen concentration) by monitoring the response of electrical conductivity to the change of $P_{O_2}$.[49,50] It is found that the thermodynamic and kinetic properties can be readily tailored by tuning the Co content in LS_MFC$_x$A. **Figure 3b** illustrates an opposite correlation of the thermodynamic (Δδ) and kinetics ($k_{ex}$) properties with the Co content, where more Co decreases the kinetics but increases Δδ. In this series, LS_MFC$_{0.4}$A likely achieved a balance between the thermodynamic and kinetics properties, delivering the highest H$_2$ yield within the given duration. The oxygen uptake and release tests further confirm the superior oxygen incorporating kinetics of LS_MFC$_x$A with a lower Co content (**Figure S10**, Supporting information). Additionally, LS_MFC$_x$A with a high Co content ($x > 0.52$) showed inferior structural stability, where the formation of secondary RP phase (**Figure 2c** and **d**) is unfavorable for the H$_2$ production. Therefore, we focused on the optimal composition LS_MFC$_{0.4}$A to further investigate its STCH performance.

**Table 2.** Summary of STCH results of LS_MFC$_{0.4}$A under different testing conditions. The reduction was conducted at 1350 °C in N$_2$ and oxidation was performed in 40% steam with N$_2$ for all cases.

| Condition # | Flow Rate (sccm) | $T_{ox}$ (°C) | $t_{Re}$ (min) | $t_{Ox}$ (min) | Cumulative H$_2$ (μmol/g) |
|---|---|---|---|---|---|
| C1 | 100 | 800 | 45 | 15 | 49 |
| C2 | 100 | 1000 | 45 | 15 | 194 |
| C3 | 100 | 1100 | 45 | 15 | 219 |
| C4 | 100 | 1100 | 30 | 30 | 270 |
| C5 | 100 | 1100 | 30 | 60 | 351 |
| C6 | 200 | 1100 | 30 | 30 | 395 |

The hydrogen production is influenced by the testing conditions, such as the gas flow rate, reaction temperature ($T_{Re}$ and $T_{Ox}$) and reaction time ($t_{Re}$ and $t_{Ox}$). Therefore, the effects of these factors on the H$_2$ yield of LS_MFC$_{0.4}$A were systematically investigated. The reduction temperature was fixed at 1350 °C. **Table 2** summarizes the cumulative H$_2$ production in the stabilized cycle under different conditions. A larger temperature swing by reducing the oxidation temperature (C1, C2 and C3) reduced the H$_2$ yield in the 1-hour cycle, although the larger temperature swing is supposed to thermodynamically favor the H$_2$ production. We believe that the oxidative water splitting step is primarily limited by the kinetics. The elevated oxidation temperature of (C3) rendered H$_2$ production four times higher than the lower oxidation temperature (C1), as the former promoted the reaction kinetics. Furthermore, when $T_{Ox}$ was higher than 1200 °C, the alumina tube produced a substantial H$_2$ background level arising from its catalytic water thermolysis; hence, we did not further investigate higher oxidation temperatures. The extended oxidation time enhanced the H$_2$ yield, confirming the kinetically limited oxidative water splitting step. Increasing the flow rate of steam also boosted the H$_2$ yield from 270 μmol g$^{-1}$ at 100 sccm to 395 μmol g$^{-1}$ at 200 sccm within 30-min oxidation time likely due to the improved mass transfer of steam and rapid removal of H$_2$ product. The highest H$_2$ yield of 395 μmol g$^{-1}$ in a short duration (30 min oxidation and 30 min reduction) for LS_MFC$_{0.4}$A is comparable to those of best reported reported STCH materials (**Table S2**, Supporting Information), indicating the great potential of CCPOs for STCH.

We investigated the cyclability of LS_MFC$_{0.4}$A under both uninterrupted and harsher interrupted conditions. **Figure 3c** exhibits the uninterrupted cycling test results (similar to most prior reports). After the initial two activation cycles, LS_MFC$_{0.4}$A maintained a stable H$_2$ yield of 270 ± 9 and 395 ± 11 μmol g$^{-1}$ at a steam flow rate of 100 and 200 sccm for the following 10 consecutive cycles, respectively. The particle morphology and EDS mapping of pristine and cycled LS_MFC$_{0.4}$A are shown in **Figure S11**. The LS_MFC$_{0.4}$A after 12 cycles did not show significant element segregation despite apparent sintered particles. Note that the seemingly inhomogeneity in EDS maps of some elements such as Sr and Al compared to other elements is owing to the different element interaction volume rather than their inhomogeneous distribution. For the practical STCH applications, the diurnal cycle of sunlight irradiation should be considered without the viable efficient heat storage technology, as the STCH materials may subject to drastic temperature swing between elevated temperatures under 8-hour strong sunlight irradiation in the day and room temperature of long-time cooling stage in the night. In stark contrast to the ideal uninterrupted cycling condition, such interrupted cycling involving startup heating and shutdown cooling is a harsh condition to the STCH application, which can induce thermal fatigue, impair the structural integrity and dramatically deteriorate the H$_2$ production cycle.[51,52] A similar challenge exists in other solar-driven hydrogen production technologies such as photovoltaic driven water electrolysis, where the startup and shutdown cycling due to the day-night cycle causes degradation of electrocatalysts.[53–56] Few research on development of STCH materials has paid attention to the cyclability under this harsh condition.[51] Therefore, the interrupted cycling stability of LS_MFC$_{0.4}$A for thermochemical hydrogen production was evaluated under the daily startup and shutdown cycling for 13 days (51 cycles) to simulate the real diurnal cycle (we note the more cycles per day can be carried out in a real reactor with faster ramping rates, in comparison with our home-made testing reactor). **Figure 3d** demonstrates that LS_MFC$_{0.4}$A had outstanding stability against the thermal fatigue and redox degradation without significant decay in H$_2$ production under this harsh condition. This is possibly attributed to the no phase transformation of the CCPO during redox cycling and a possible contribution from entropy stabilization, which make our CCPO stand out among other state-of-the-art STCH materials (Table S2). The SEM images and EDX mapping of LS_MFC$_{0.4}$A after 51 cycles displayed sintered morphology but homogeneous elemental distribution (**Figure S11c**, Supporting information).

To elucidate why Co is the most redox active in LS_MFC$_x$A and shed light on the factors governing the priority of redox behavior, we employed a new computation paradigm to conduct Monte Carlo (MC) sampling based on DFT. Four major initial perovskite structures, with space groups of $Pnma$, $R\bar{3}c$, and $Pm\bar{3}m$ as base structures, were used to generate randomly mixed 80-atom (La$_{0.75}$Sr$_{0.25}$)(Mn$_{0.25}$Fe$_{0.25}$Co$_{0.25}$Al$_{0.25}$)O$_3$ supercells (as the "seed"). Vacancy supercells were created by removing one of each equivalent oxygen from base structures before the cations were randomly mixed. It was found that $Pnma$, $R\bar{3}c$, and $Pm\bar{3}m$ structures have similar energies and configurations. Thus, only $R\bar{3}c$ results were reported here, consistent with the XRD patterns of LS_MFC$_x$A. We performed MC sampling of both the cation distribution and magnetic configurations at $T_{MC}$ = 1600 K and analyzed a number of 5600 equilibrated structures after 500 MC steps.

For bulk structures without oxygen vacancies, the nearest neighbor check was adopted to confirm if there are any B-cations' configurational preferences. The oxygen atom is surrounded by four first nearest neighbor (FNN) with two A-site atoms and two B-site atoms. In the accepted low energy (La$_{0.75}$Sr$_{0.25}$)(Mn$_{0.25}$Fe$_{0.25}$Co$_{0.25}$Al$_{0.25}$)O$_3$ 80-atom supercell, the probability of finding oxygen FNN B-site elements to be the same type of $B_1$ element is $P(B_1 - B_1) = \frac{4}{16} \times \frac{3}{15} = 0.05$. The probability to find two atoms of different elements $B_1 - B_2$ is $P(B_1 - B_2) = 2 \times \frac{4}{16} \times \frac{4}{15} = \frac{2}{15} \approx 0.133$, suggesting no configurational preference for B-site cation combinations. Therefore, the cation configuration of the fully oxidized bulk structure is similar to a random distribution without signatures of B-site short-range ordering consistent with the defined CCPO structure. By introducing a single neutral oxygen vacancy, the configurational preference is changed. For the accepted vacancy containing configurations, the B-site distribution in **Figure 4a** shows that 83% of the combinations prefer to have at least one Co at its FNN position in $R\bar{3}c$ structure, strongly deviating from the probability in a random distribution (0.44%). The possible positions of formed oxygen vacancy are illustrated in **Figure 4d** (Fe-Vac-Co) and **Figures S12–S18** (Supporting Information). A further analysis of the magnetic moments of the B-site transition metal elements (excluding the non-magnetic Al) in the accepted low energy bulk and vacancy systems provides information about how the oxygen vacancy changes the cation valence states. The magnetic moment magnitude is directly correlated with the oxidation state of the magnetic transition metal. The statistical distribution plotted in **Figure 4b** shows the respective magnetic moment magnitude changes of Mn, Fe, and Co in bulk and vacancy configurations. Before the vacancy is formed (*i.e.*, bulk), the median magnetic moment magnitudes for Mn, Fe, and Co are 3.0, 4.0, and 2.9, corresponding to oxidation states of $Mn^{4+}$, $Fe^{3+}$, and $Co^{3+}$, respectively. After introducing a single oxygen vacancy, Co is the element that has the greatest magnetic moment magnitude change ($Co^{3+} \rightarrow Co^{2+}$) and broadest distribution. The magnetic moment distribution of Mn is slightly broadened. The magnetic moment distribution of Fe is almost unchanged. Therefore, the $Co^{3+}/Co^{2+}$ redox pair is the most active in response to the formation of oxygen vacancies in the LS_MFC$_x$A systems, possibly followed by $Mn^{4+}/Mn^{3+}$.

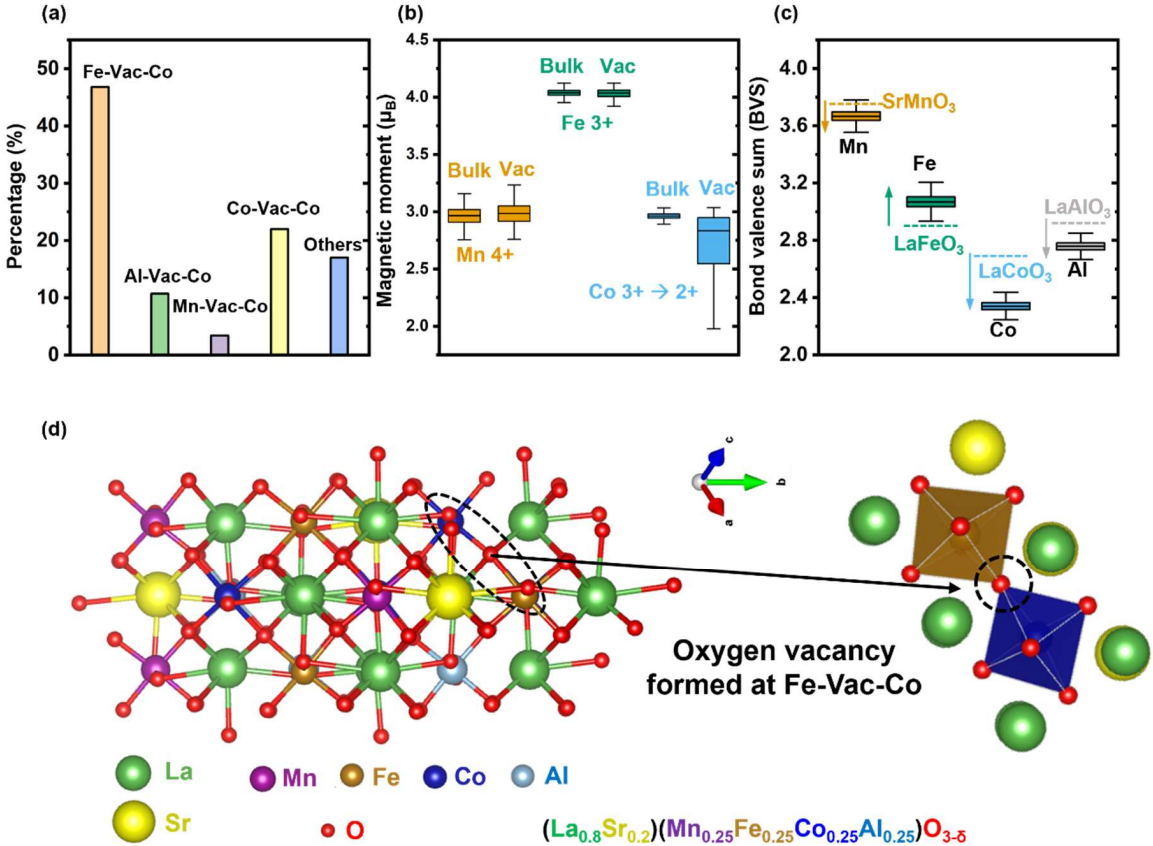

**Figure 4. a)** Vacancy (Vac) B-site first nearest neighbors (FNN) distributions show dominant Co preference over other B-site elements. **b)** Local magnetic moment evolution between bulk and vacancy configurations for 3*d* orbitals of Mn, Fe and Co from saved low energy bulk structures (~2600 configurations) and vacancy structures (~3000 configurations) shows that Co goes through the largest magnitude change when a single neutral oxygen vacancy is formed. The bulk structure represents the supercell structure before introducing an oxygen vacancy. **c)** Calculated bond valence sum (BVS) for Mn, Fe, Co, and Al octahedra in $(La_{0.75}Sr_{0.25})(Mn_{0.25}Fe_{0.25}Co_{0.25}Al_{0.25})O_3$ structure (solid lines for median BVS values and box plots). The dash lines present the BVS values of $SrMnO_3$, $LaFeO_3$, $LaCoO_3$, and $LaAlO_3$ simple perovskite references based on our DFT calculations. **d)** Representative of the simulation cell containing oxygen vacancy with Fe and Co as FNN of the position with the largest percentage where an oxygen vacancy is formed.

A possible reason for the Co preference in vacancy configurations is the Co-O bond stretching effect. Since neutral vacancy formation causes volume expansion in many oxides, and a tensile stress (bond stretching) in the bulk will lower the vacancy formation energy.[57–59] In the CCPOs, local bond stretching may create more space to favor oxygen vacancy formation. To capture the local bond strength, the bond valence sum (BVS) descriptor was analyzed (see details in Supporting Information).[60] The BVS for each B-site cation in a perfect structure is close to its oxidation number. A lower (higher) value is considered as an elongated (compressed) bond with weaker (stronger) bond strength. For example, using the tabulated parameters (developed by fitting experiments[61] and listed in **Table S3**), the BVS for $Mn^{4+}$ in $SrMnO_3$, $Fe^{3+}$ in $LaFeO_3$, and $Co^{3+}$ in $LaCoO_3$, and $Al^{3+}$ in $LaAlO_3$ (dash lines in **Figure 4c**) are very close to 4.0, 3.0, 2.9, and 3.0 for their DFT computed relaxed structures, respectively. The DFT calculated BVS of ternary perovskites (*e.g.*, $SrMnO_3$, $LaFeO_3$, $LaCoO_3$, and $LaAlO_3$) serves as the references to compare with that of the bulk $(La_{0.75}Sr_{0.25})(Mn_{0.25}Fe_{0.25}Co_{0.25}Al_{0.25})O_3$ (solid lines and box plots in **Figure 4c**). The distribution of BVS for Co in $(La_{0.75}Sr_{0.25})(Mn_{0.25}Fe_{0.25}Co_{0.25}Al_{0.25})O_3$ is greatly lowered compared to that in the DFT reference structure ($LaCoO_3$), indicating a distortion induced bond elongation and weakening effect. The distributions of BVS for Mn and Al in $(La_{0.75}Sr_{0.25})(Mn_{0.25}Fe_{0.25}Co_{0.25}Al_{0.25})O_3$ are slightly lower than those in their DFT-relaxed ternary perovskite references, while the BVS for Fe in $(La_{0.75}Sr_{0.25})(Mn_{0.25}Fe_{0.25}Co_{0.25}Al_{0.25})O_3$ is increased compared to that in $LaFeO_3$ reference. These results demonstrate greatly weakened Co-O bond, slightly weakened Mn-O and Al-O bonds and strengthened Fe-O bonds in CCPO.

We further evaluated the deviation of BVS on each cation in $(La_{0.75}Sr_{0.25})(Mn_{0.25}Fe_{0.25}Co_{0.25}Al_{0.25})O_3$ from its oxidation number, and the root-mean-square value of this deviation for all the cations (*i.e.*, the global stability index $G$, as detailed in Supporting Information). All Sr and Fe ions are under compression and all La, Mn, Co, Al are under tension, while Co shows the largest negative deviation from its perfect BVS. The calculated global instability index $G$ for saved bulk structures ranges from 0.217 to 0.233 vu, indicating strongly strained structures. According to the BVS theory, structures with $G > 0.2$ are rare.[62] Hence, we propose that the Co preference in vacancy configurations is due to Co-O bond stretching effect. Although $Mn^{4+}$ may be more reducible than $Co^{3+}$ based on the classical charges, the Co-O bond stretching outweighs the tendency to generate vacancies near $Mn^{4+}$, resulting the tendency to form oxygen vacancies adjacent to Co at its FNN position.

In addition to the proven excellent thermal stability and intrinsic surface oxygen exchange kinetics, two thermodynamic properties, enthalpy and entropy of reduction, are important factors determining the performance of STCH materials.[67–69] Therefore, we carried out the measurements of the oxygen non-stoichiometry over a range of temperatures and $P_{O_2}$ values for LS_MFC$_{0.4}$A using a widely reported TGA protocol (see details in **Figure S19**, Supporting Information).[19,66,70–72] **Figure 5a** shows the oxygen non-stoichiometry of LS_MFC$_{0.4}$A as a function of temperature measured by continuous ($P_{O_2} \geq 0.028$ atm) and stepwise ($P_{O_2} \leq 1.7 \times 10^{-3}$ atm) heating under the given $P_{O_2}$ values. Following the van't Hoff method,[19,63,73] the enthalpy and entropy of reduction can be correlated with temperature and $P_{O_2}$ according to the following equation:

$$R\ln(P_{O_2})^{\frac{1}{2}} = -\frac{\Delta_{red}H(\delta)}{T} + \Delta_{red}S(\delta) \quad (1)$$

where $R$ is the universal gas constant, and $P_{O_2}$ is equal to the oxygen partial pressure referenced to the standard gas pressure (1 atm). $T$ is the temperature in Kelvin. $\Delta_{red}H(\delta)$ and $\Delta_{red}S(\delta)$ are the standard enthalpy and entropy of reduction, respectively, which are dependent on δ and defined on a per mole of oxygen basis but have negligible dependence on temperature.

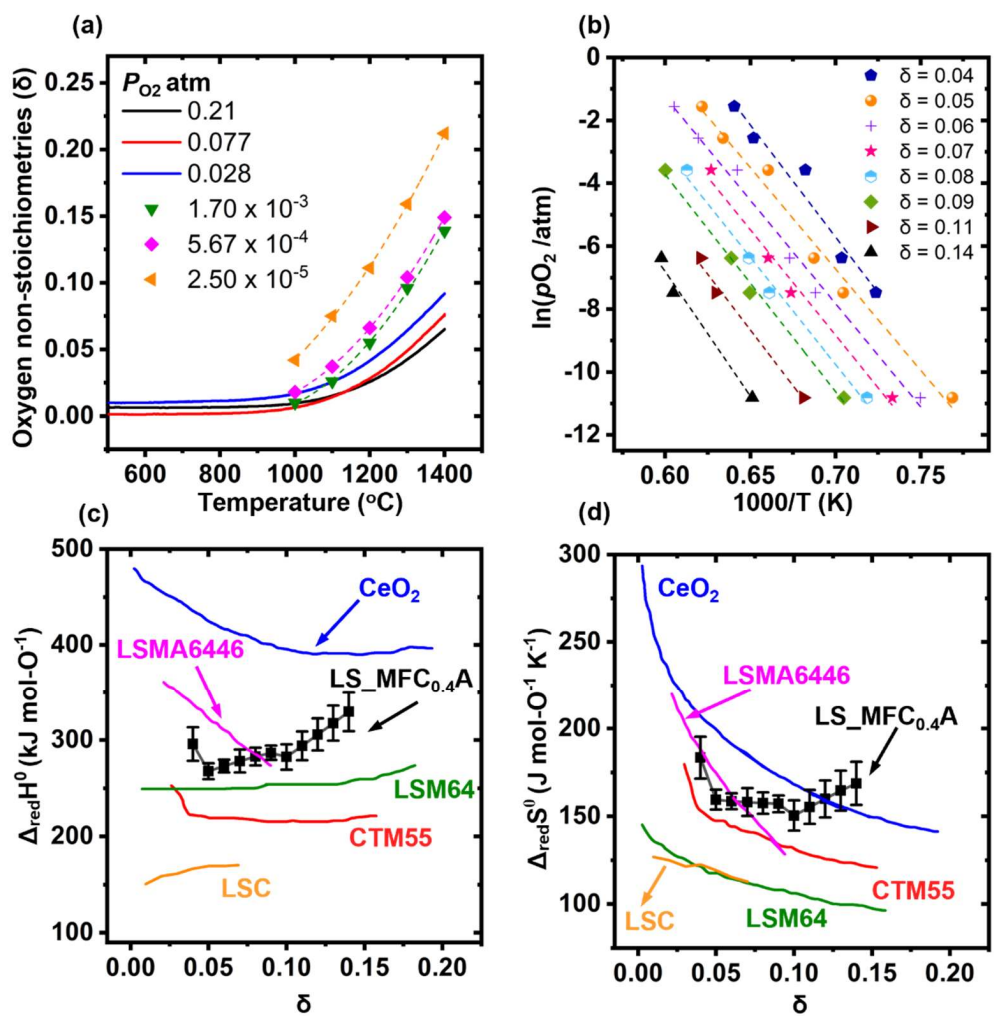

**Figure 5. a)** Oxygen non-stoichiometry of LS_MFC$_{0.4}$A as a function of temperature as measured by continuous (solid lines) and stepwise (symbols) heating under the given $P_{O_2}$ values. **b)** Arrhenius representation of $\ln(P_{O_2})$ vs. 1000/$T$ for extraction of thermodynamic properties (e.g., enthalpy and entropy of reduction) by the van't Hoff method at typical fixed δ values as indicated. Symbols and dash lines in the profile represent the measured and linearly fitted results, respectively. **c)** Standard enthalpy and **d)** entropy of reduction calculated for LS_MFC$_{0.4}$A and comparison to those of some reported STCH and redox oxide materials. The error bars of LS_MFC$_{0.4}$A came from the error of its linear fitting. The literature data for CeO$_2$,[63] La$_{0.6}$Sr$_{0.4}$Mn$_{0.4}$Al$_{0.6}$O$_3$ (LSMA6446),[64] cubic CaMn$_{0.5}$Ti$_{0.5}$O$_3$ (CTM55),[19] La$_{0.6}$Sr$_{0.4}$MnO$_3$ (LSM64),[65] and La$_{0.8}$Sr$_{0.2}$CoO$_3$ (LSC)[66] is presented for comparison.

An Arrhenius plot of $\ln(P_{O_2})$ vs. $1000/T$ at each specified δ value with linear fitting can be obtained (**Figure 5b**), from which $\Delta_{red}H(\delta)$ and $\Delta_{red}S(\delta)$ can be extracted from the slope and intercept, respectively. In this way, the enthalpy and entropy of LS_MFC$_{0.4}$A as a function of δ are obtained (**Figure 5c** and **5d**) and compared with those of some reported STCH and redox oxide materials such as CeO$_2$,[63] cubic Ca(Mn$_{0.5}$Ti$_{0.5}$)O$_{3-δ}$ (CTM55),[19] La$_{0.6}$Sr$_{0.4}$MnO$_{3-δ}$ (LSM64),[65] La$_{0.6}$Sr$_{0.4}$Mn$_{0.4}$Al$_{0.6}$O$_{3-δ}$ (LSMA6446),[64] and La$_{0.8}$Sr$_{0.2}$CoO$_3$ (LSC).[66]

Previous experimental and theoretical studies suggested that a combination of intermediate enthalpy and large entropy is favorable for STCH from water splitting.[19,64,72,74] For perovskite materials, the enthalpy of reduction can be tuned by the dopants,[75] and is desirably constrained to a certain moderate range (neither too high nor too low), which may follow a Sabatier principle. A low enthalpy of reduction usually accompanies a low entropy of reduction, rendering favorable reduction extents (large Δδ) but unfavorable oxidative water splitting.[74,76] In contrast, a high enthalpy of reduction enables a high energy penalty for thermal reduction. LS_MFC$_{0.4}$A shows enthalpy and entropy values comparable to LSMA6446 and CTM55 which have been identified to have prominent hydrogen production with an appropriate combination of moderate enthalpy and relatively large entropy.[19,64] Compared with ceria, LS_MFC$_{0.4}$A possess a much lower enthalpy but similar entropy of reduction when Δδ is more than 0.1, suggesting its potential for high H$_2$ yield under a lower reduction temperature. As shown in the TPR results (**Figure 2**), the Δδ value of LS_MFC$_x$A increases with $x$, possibly due to the decreased enthalpy of reduction upon the growing content of Co. For example, LSC material shows a small enthalpy (< 200 kJ mol$^{-1}$) and entropy simultaneously, suggesting that it tends to have a large Δδ but poor oxidative water splitting capability. This is in good agreement with our TGA and STCH results for LSC. LS_MFC$_{0.4}$A displays the highest hydrogen production in this LS_MFC$_x$A series. This finding suggests that while maintaining a moderate enthalpy of reduction, multiple cations in LS_MFC$_x$A may introduce extra configurational entropy to fulfill the needs of oxygen vacancy formation and electronic and ionic configurational entropy change.[77,78]

## 3. Conclusions

In this study, we demonstrated a new class of medium- to high-entropy CCPOs, (La$_{0.8}$Sr$_{0.2}$)(Mn$_{(1-x)/3}$Fe$_{(1-x)/3}$Co$_x$Al$_{(1-x)/3}$)O$_3$ with tunable thermodynamic and kinetics properties for two-step thermochemical water splitting. The reversible extent of reduction (Δδ) increases with the rising Co content ($x$), whereas the intrinsic kinetics (oxygen surface exchange coefficient) shows a decreased trend with the increasing $x$. (La$_{0.8}$Sr$_{0.2}$)(Mn$_{0.2}$Fe$_{0.2}$Co$_{0.4}$Al$_{0.2}$)O$_3$ ($x = 0.4$) or LS_MFC$_{0.4}$A exhibits an optimal balance between intrinsic thermodynamics and kinetics, as well as exceptional structural stability during the redox cycling and a favorable combination of moderate reduction enthalpy (268–329 kJ (mol-O)$^{-1}$) and high entropy (150–180 J (mol-O)$^{-1}$ K$^{-1}$). These merits of LS_MFC$_{0.4}$A enable a maximum hydrogen yield of 395 ± 11 µmol g$^{-1}$ within a short 1 h duration (30 min reduction and 30 min oxidation) at an optimized STCH conditions as well as remarkable STCH cycling durability. The preferred redox of Co over other B-site metals is revealed by the TPR and *in-situ* XPS. A parallel MC/DFT computation demonstrates that an oxygen vacancy prefers to form in the vicinity of the position with at least one Co at its first nearest neighbor. An analysis of the magnetic moments of the B-site metals confirms that the Co redox valence change is the most active in response to the formation of oxygen vacancies. The bond valence sum results demonstrate greatly weakened Co-O bond, slightly weakened Mn-O and Al-O bonds and strengthened Fe-O bonds. The Co-O bond stretching effect rationalizes the dependence of Δδ on Co content and its dominance in the redox of LS_MFC$_x$A.

In general, we showed the importance to achieve a trade-off of thermodynamic and kinetic properties for optimized STCH performance. This study further exemplifies that non-equimolar compositions outperform their higher-entropy equimolar counterparts in this two-step STCH application. Thus, while an entropy effect with the multiple cations may stabilize oxygen vacancy formation without phase transformation, maximizing configurational entropy is not needed. Instead, we should use non-equimolar compositional design to balance the thermodynamic and kinetic properties to achieve better STCH performance, which represent a new design strategy in general.

This work provides a new pathway to explore CCPOs with tunable redox, thermodynamic and kinetics properties for STCH and chemical looping and a new computation paradigm to predict the redox behavior.

## 4. Experimental section

**Material synthesis:** The CCPO samples of (La$_{0.8}$Sr$_{0.2}$)(Mn$_{(1-x)/3}$Fe$_{(1-x)/3}$Co$_x$Al$_{(1-x)/3}$)O$_3$ (denoted as "LS_MFC$_x$A") were prepared by solid-state reactions. The starting powders, La$_2$O$_3$ (99.99%), SrCO$_3$(99.9%), MnO$_2$(99.5%), Fe$_2$O$_3$ (99.99%), Co$_3$O$_4$ (99.9%) and Al$_2$O$_3$ (99.99%), were purchased from Alfa Aesar. The precursor powders were mixed based on the calculated stoichiometry and placed in a poly(methyl methacrylate) high-energy ball mill (HEBM) vial with endcaps and milling balls made by tungsten carbide. The vials were dry milled for 100 min (SPEX 8000D, SPEX SamplePrep, USA). Then, the mixed powder was annealed in air at 1300 °C for 10 hours to form the single perovskite phase. The synthesized powder was ground by pestle and mortar and further annealed in air at 1350 °C for another 10 hours to improve the homogeneity.

Additional experimental details of materials characterization, measurements, in-situ experiments, and modeling can be found in Suppl. Mater. Section A.

**Acknowledgement**

This work is supported by the U.S. Department of Energy (DOE), Office of Energy Efficiency and Renewable Energy (EERE), under the Agreement Number DE-EE0008839, managed by the Hydrogen and Fuel Cell Technologies Office in the Fiscal Year 2019 H2@SCALE program. Sandia National Laboratories is a multi-mission laboratory managed and operated by National Technology and Engineering Solutions of Sandia, LLC., a wholly owned subsidiary of Honeywell International, Inc., for the U.S. Department of Energy's National Nuclear Security Administration under contract DE-NA0003525. The Alliance for Sustainable Energy, LLC, operates and manages the National Renewable Energy Laboratory (NREL) for DOE under contract no. DE-AC36-08GO28308. The research was performed using high-performance computing resources sponsored by DOE-EERE and located at NREL.